\documentclass[twocolumn,showpacs,preprintnumbers,amsmath,amssymb]{revtex4}

\usepackage{graphicx}
\usepackage{dcolumn}
\usepackage{bm}

\begin{document}

\preprint{Cerchez et al.}

\title{Effect of edge transmission and elastic scattering on the resistance of magnetic barriers}

\author{M. Cerchez}\author{S. Hugger}\author{T. Heinzel}
\email{thomas.heinzel@uni-duesseldorf.de}
\affiliation{Heinrich-Heine-Universit\"at, Universit\"atsstr.1,
40225 D\"usseldorf, Germany}
\author{N. Schulz}
\affiliation{Fraunhofer Institut f\"ur Angewandte
Festk\"orperphysik, Tullastr. 72, 79108 Freiburg, Germany}
\date{\today}

\begin{abstract}
Strong magnetic barriers are defined in two-dimensional electron
gases by magnetizing dysprosium ferromagnetic platelets on top of a
Ga[Al]As heterostructure. A small resistance across the barrier is
observed even deep inside the closed regime. We have used
semiclassical simulations to explain this behavior quantitatively in
terms of a combined effect of elastic electron scattering inside the
barrier region and E x B drift at the intersection of the magnetic
barrier with the edge of the Hall bar.
\end{abstract}

\maketitle

\section{\label{sec:1}INTRODUCTION}

A two-dimensional electron gas (2DEG) reacts sensitively to
perpendicular magnetic fields. Making the magnetic field
inhomogeneous opens the door to a wide variety of fascinating
effects and applications, ranging from magnetic superlattices
\cite{Ye1995} and magnetic wave\-guides \cite{Nogaret2003} to Hall
sensors for magnetic nanostructures.
\cite{Peeters1998,Novoselov2002} One particularly simple magnetic
field structure is the \emph{magnetic barrier}, namely a
perpendicular magnetic field configuration strongly localized along
one in-plane direction and homogeneous in the second one.
\cite{Peeters1993,Matulis1994} In a classical picture, magnetic
barriers can be considered as selective transmitters that filter the
electrons according to their angle of incidence. \cite{Peeters1993}
In a ballistic sample without edges, one would therefore expect that
above a critical barrier height the barrier \emph{closes}, all
electrons are reflected and the resistance approaches infinity.
Recently, magnetic barriers have received renewed interest due to
their potential applications as tunable spin filters and detectors,
both of which are highly desirable spintronics devices.
\cite{Majumdar1996,Guo2000,Papp2001a,Papp2001b,Xu2001,Guo2002,Jiang2002,
Zhai2005,Zhai2006} These theoretical works suggest that a high
degree of spin polarization may be achievable with magnetic barriers
in suitable materials.

Considering the elementary character and the simplicity of a
magnetic barrier, it is surprising that only a few transport
experiments on such structures have been reported. In
\cite{Leadbeater1995}, a magnetic barrier with a square profile has
been experimentally realized in a highly sophisticated sample,
namely a 2DEG containing a graded step. It was observed that even
for strong magnetic fields, the barrier resistance remains finite.
The results of these experiments have been subsequently interpreted
within a classical model \cite{Ibrahim1997}, which shows that
$E\times B$ drift effects at the edge of the 2DEG, as well as
elastic scattering, limit the resistance to finite values.

In all other experiments reported so far except ref.
\cite{Leadbeater1995}, the magnetic barrier has been generated in
conventional Ga[Al]As heterostructures by magnetizing a
ferromagnetic platelet, located on top of the sample, by an in-plane
magnetic field. \cite{Monzon1997,Kubrak2000,Vancura2000,Kubrak2001,
Gallagher2001,Hong2002} In such a setup, the magnetic barrier
originates from the z-component of the stray field of the
ferromagnet, see Fig. 1. This experimental implementation is also
the basis for a significant fraction of the theoretical studies.
\cite{Lu2002,Xu2005,Zhai2005,Zhai2006}\\

For an experimental implementation of the theoretical concepts, a
detailed and quantitative understanding of the measured transmission
properties of tunable magnetic barriers is needed. Previous studies
have already shown that both edge transmission and scattering in the
barrier region are relevant. \cite{Kubrak2001,Hong2002} Here, we
build on these results and discuss in detail how the resistance of
tunable magnetic barriers depends upon the $E \times B$ drift at the
edges, on the elastic scattering and on thermal smearing. In order
to magnify these influences, we have prepared our ferromagnetic
films from dysprosium which has a particularly large bulk saturation
magnetization of $\mu_0 M = 3.75\,\mathrm{T}$.
\cite{Behrendt1958,Stepankin1995} This allows us to drive the
barriers well into the closed regime, where the transport through
the structure is exclusively determined by the effects of interest
here. In addition, a top gate was used to tune the electron density.

These measurements are interpreted in a semi-classical picture based
on the billiard model for ballistic conductors.
\cite{Buttiker1986,Beenakker1989} We find that (i) the combination
of both $E \times B$ type edge drifts and elastic scattering in the
barrier determines the barrier resistance, (ii) reasonable
assumptions regarding the distribution of scattering angles for the
elastic electron scattering lead to excellent agreement of the
experimental data with the model, and (iii) thermal smearing has a
marginal influence at liquid helium temperatures.

The outline of the paper is as follows: in Section II, we describe
the sample preparation, the experimental setup and the measurement
methodology. The experimental results are presented in Section III,
while the semi-classical model and its application to our
measurements is the topic of Section IV. The paper concludes with a
summary and a discussion (Section V).

\section{\label{sec:2}SAMPLE PREPARATION AND EXPERIMENTAL SETUP}

\begin{figure}
\includegraphics[scale=0.9]{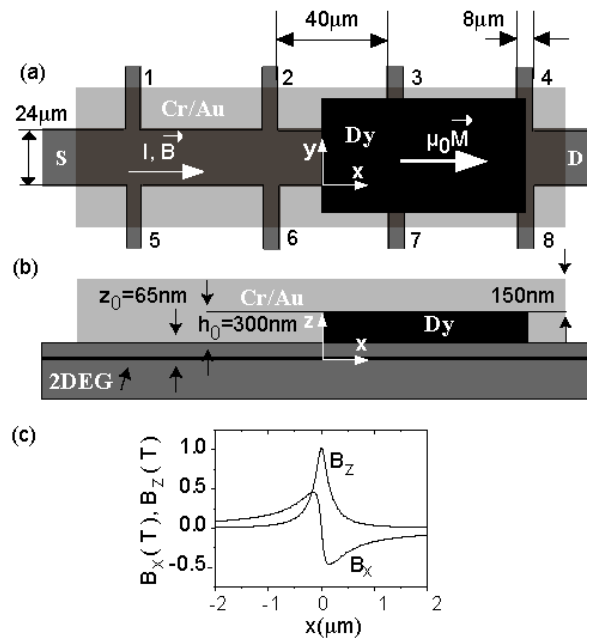}
\caption{(a) Top view of the sample: a dysprosium platelet is placed
on top of a Hall bar with source and drain contacts (left and right)
and voltage probes 1 to 8. The structure is covered by a homogeneous
Cr/Au gate. The parallel magnetic field is applied in the
x-direction. (b) Cross section of the sample in the x-z plane. (c)
At the edges of the Dy film, the fringe field in the z-direction
$B_z(x)$ is highly localized in the x-direction. Also shown is the
x-component of the fringe field. } \label{mb1}
\end{figure}

A commercially available $\mathrm{GaAs/Al_xGa_{1-x}As}$ -
heterostructure \cite{IntelliEpi} with a 2DEG $65\,\mathrm{nm}$
below the surface was laterally patterned by using optical
lithography and subsequent processing steps. A Hall bar geometry
(Fig. \ref{mb1}) was defined by wet chemical etching. Au/Ge ohmic
contacts were defined at source and drain contacts and at the
voltage probes 1 to 8. A dysprosium (Dy) platelet with a thickness
of $h_0=300\,\mathrm{nm}$ was defined at the heterostructure surface
by Dy thermal evaporation at a base pressure of $2\times
10^{-6}\,\mathrm{mbar}$. A Cr/Au gate layer of $150\,\mathrm{nm}$
thickness was deposited on top to prevent the Dy from oxidizing
under ambient conditions and to allow the carrier density to be
tuned. Six samples were measured, and all showed qualitatively
identical behavior. Here, we discuss data taken from one
representative sample.

The samples were inserted in a liquid helium cryostat with a
variable temperature insert that permits variation of the
temperature between $2\,\mathrm{K}$ and room temperature. The sample
stage is equipped with a rotatable sample holder, such that the
magnetic field could be oriented within the x-z plane with an
accuracy better than $0.1$ degrees.

The carrier densities and the electron mobility were determined from
conventional four-probe  measurements of the components of the
resistance tensor, $R_{xx}$ and $R_{xy}$ in perpendicular magnetic
fields. The ungated electron density is $2.3 \times
10^{15}\,\mathrm{m^{-2}}$, and the mobility at a temperature of
$2\,\mathrm{K}$ is $29.0\,\mathrm{m^2/Vs}$, corresponding to a Drude
scattering time of $\tau_D = 10.2\,\mathrm{ps}$ and an elastic mean
free path of $2.2\,\mathrm{\mu m}$. The quantum scattering time was
determined from the envelope of the Shubnikov - de Haas oscillations
\cite{Ando1982} as $\tau_q =1.05\,\mathrm{ps}$. The vanishing of the
Hall voltage was furthermore used to detect the parallel magnetic
field configuration. At the maximum magnetic field $B$ of
$8\,\mathrm{T}$ used in our experiments, we estimate that the
maximum external perpendicular magnetic field component is below
$5\,\mathrm{mT}$.

Strong parallel magnetic fields are well known to affect the
transport properties of 2DEGs by modifying the density of states and
the interactions. \cite{Lee1985} In addition, the electron effective
mass becomes slightly anisotropic. \cite{Bhattacharya1982} These
effects show up as a weak and approximately parabolic
magnetoresistivity.

\begin{figure}
\includegraphics[scale=0.45]{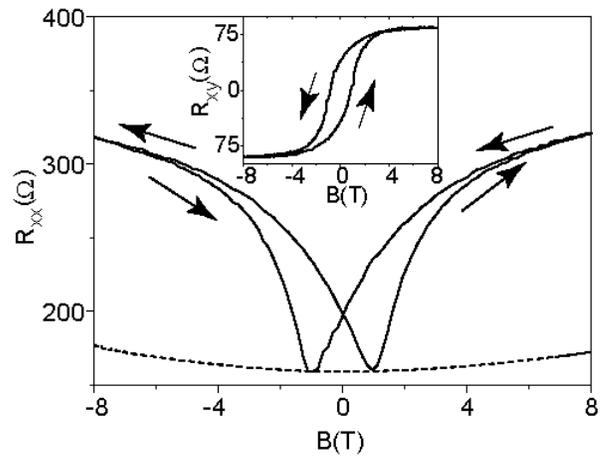}
\caption{Resistance of the magnetic barrier as a function of the
magnetic field (full lines), as measured between contacts 2 and 3.
The dashed line shows the magnetoresistance of the 2DEG, measured
between contacts 1 and 2. Inset: Hall resistance of the magnetic
barrier (the voltage was collected between contacts 4 and 8). The
arrows denote the scan directions.} \label{mb2}
\end{figure}

Increasing B also magnetizes the Dy film along the x-direction. The
z-component of the fringe field at the 2DEG is strongly localized at
the edge of the Dy in the x-direction and forms the magnetic
barrier. \cite{Johnson1997,Monzon1997} The x- and z- components of
the fringe field are shown in Fig. \ref{mb1}(c) for the literature
value of the saturation magnetization of Dy. Assuming that the
fringe field follows the corresponding analytic expressions
\cite{Ibrahim1997,Vancura2000}, $B_z(B,x)$ is given by

\begin{equation}
B_z(B,x)=-\frac{\mu_0M(B)}{4\pi}\ln{\left(\frac{x^2+z_0^2}{x^2+(z_0+h_0)^2}\right)}
\label{eq1}
\end{equation}

where $z_0$ is the distance of the 2DEG from the surface and $h_0$
denotes the thickness of the Dy film, see Fig. \ref{mb1}(c). This
relation neglects the second magnetic barrier residing at contacts 4
and 8, which is justified since it is sufficiently far away from the
region probed between contacts 2 and 3. Edge roughness of the
magnetic film may also lead to deviations from eq. \eqref{eq1}. We
characterized the Dy edge by atomic force microscope measurements
and found an edge roughness (single standard deviation) of
$35\,\mathrm{nm}$, which is smeared out to a large extent at the
2DEG. We therefore neglect the edge roughness in the following. The
magnetization in the x-direction as a function of $B$ is denoted by
$M(B)$, which can be estimated from Hall measurements on a magnetic
barrier \cite{Monzon1997} as described below. The x-component of the
fringe field has a much smaller effect on the 2DEG and is moreover
small compared with the B field required to establish saturation
(i.e. $B_{sat}\approx 4.5\,\mathrm{T}$, see the inset in Fig.
\ref{mb2}). It is therefore neglected in the following.

A current of $100\,\mathrm{nA}$ at a frequency of
$13.6\,\mathrm{Hz}$ is passed from source to drain. The barrier
resistance is obtained from the voltage measured between contacts 2
and 3 (Fig. 1) with a lock-in amplifier. The Hall voltage measured
between contacts 4 and 8 is used to determine $M(B)$. We assume in
the discussion below that the magnetic barriers at both edges of the
Dy platelet differ only by their sign.

\section{\label{sec:3}EXPERIMENTAL RESULTS}

Figure 2 shows a typical magnetoresistance trace $R_{xx}(B)$ of our
samples, measured on sample A. The traces are hysteretic, reflecting
the magnetization characteristics of the Dy film (inset). At the
coercive magnetic fields at $\pm 0.9\,\mathrm{T}$, $R_{xx}(B)$ has
minima which equal to high accuracy the magnetoresistance outside
the Dy film measured over an identical distance. This shows that the
micromagnetic structure in the Dy film which becomes most relevant
around the coercive magnetic field \cite{Vancura2000}has no
noticeable effect in our experiments. $R_{xx}(B)$ increases as one
goes away from the coercive field, but neither saturates nor
approaches infinity, even well above the saturation magnetic field.
The slope $dR_{xx}/dB$ above $B_{sat}$ is only partly due to the
parabolic background. As discussed in more detail below, if our
barrier were fully ballistic and without edges, it would close at
$B=0.7\,\mathrm{T}$ away from the minimum of $R_{xx}$. Thus, in 88\%
of the magnetic field interval scanned, the transmission is governed
by the edge and scattering effects of interest. Note that the
$R_{xx}(B)$ traces are slightly asymmetric around their minimum. We
attribute this effect to the proximity of the voltage probes
($17\,\mathrm{\mu m}$) to the magnetic barrier, over which the
electrons ejected from the barrier may not yet form a Fermi sphere,
even though the probes are about 8 elastic mean free paths away from
the barrier. Similar effects have been observed by Leadbeater et al.
\cite{Leadbeater1995} and subsequently been explained in detail by
Ibrahim et al. \cite{Ibrahim1997}.

The Hall resistance of the magnetic barrier (inset in Fig.
\ref{mb1}) measures $B_z(B,x)$, averaged over the spatial extension
$L$ in the x-direction of the Hall probe contacts 4 and 8, i.e. over
$8\,\mathrm{\mu m}$, according to

\begin{equation}
R_{xy}(B)=-\frac{\alpha}{L ne} \int\limits_{-L/2}^{L/2}B_z(B,x)dx
\label{eq2}
\end{equation}

Here, $\alpha$ represents a Hall factor which may deviate from 1,
depending on the sample geometry and the mean free path.
\cite{Reijniers1998,Reijniers2000,Liu1998} For our structure,
$\alpha\approx 1$ is expected \cite{Peeters1998}, and we have no
reason to assume otherwise, in contrast to the findings reported in
Ref. \cite{Vancura2000}. Here, we have assumed that the magnetic
barrier on top of our Hall cross is adequately described by the
ballistic model, even though the mean free path is smaller that the
width of the voltage probes. Since, however, the FWHM of our barrier
in the closed regime is no larger than $300\,\mathrm{nm}$ and thus
much smaller than the mean free path, and the maximum magnetic field
multiplied by the electron mobility $B_{max}\times \mu \gg 1$ for
our system in the closed regime, the diffusive model developed in
Ref. \cite{Cornelissens2002} does not strictly apply as well,
\cite{Reijniers2000} while to the best of our knowledge, a model for
the intermediate regime is not available. For $\alpha = 1$, we
determine from the measured $R_{xy}(B)$ a saturation magnetization
of $M_s=1.9\,\mathrm{T}$ for our Dy films. This is significantly
below the literature value for bulk dysprosium. We attribute this
reduction to the embedding of oxygen into the Dy film during
metallization, as well as to the granularity of the
film.\cite{Stepankin1995} This interpretation is supported by our
observation of $\mu_0 M_s$ dropping over time in samples where the
Dy films is not covered by a Cr/Au layer.

\begin{figure}
\includegraphics[scale=0.4]{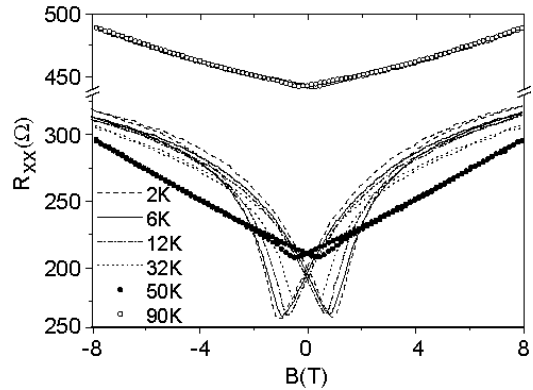}
\caption{Temperature dependence of the magnetic barrier
magnetoresistance.} \label{mb3}
\end{figure}

Figure \ref{mb3} reproduces the behavior of the barrier
magnetoresistance as the temperature is changed. As the temperature
is increased, the magnetic barrier resistance is reduced. At the
same time, the shape of the barrier becomes more nearly triangular
and the hysteresis decreases. Simultaneously, the resistance minima
increase and are shifted to smaller absolute values of $B$. As will
be explained in more detail below, this behavior can be understood
in terms of a combination of increased scattering and a reduced
coercivity of the Dy film as the temperature is increased, while the
thermal smearing of the Fermi function plays a marginal role. A
slight hysteresis is observed even above the literature value for
the Curie temperature of bulk Dy, $T_C=85\,\mathrm{K}$.
\cite{Behrendt1958} The enhancement of the Curie temperature is a
second indication of
crystal imperfections in our Dy films. \cite{Stepankin1995}\\

\section{\label{sec:4}SIMULATION AND INTERPRETATION OF THE EXPERIMENTS}

Our numerical approximation of the magnetic barrier resistance is
exemplified using the data of Fig. \ref{mb2}. The analysis is based
upon the billiard model for quasi-ballistic conductors
\cite{Beenakker1989} and the Landauer-B\"uttiker formalism
\cite{Buttiker1986}. Electrons are injected into the barrier region
starting from a fixed x-position $10\,\mathrm{\mu m}$ to the left of
the barrier at random positions in the y-direction across the Hall
bar of width $W=24\,\mathrm{\mu m}$. They start out with the Fermi
velocity, while their directions are arbitrarily distributed over
all angles with positive x-component. We solve the differential
equations describing the classical motion of the electrons to obtain
their trajectories until either it is rejected by the barrier and
passes the starting line in the opposite direction, or until it
travels through the barrier and reaches the x position
$10\,\mathrm{\mu m}$ to its right. At this distance, $B_z(x)$ is
negligible. Injecting the electrons at larger distances does not
modify the results. The edges of the Hall bar have been incorporated
by constant, reflective electric fields in the regions
$|y|>12\,\mathrm{\mu m}$. Strong electric edge fields generate
specular reflection. The magnetic barrier is incorporated as given
by eq. \eqref{eq1} in the x-direction and homogeneous in the
y-direction. We have introduced scattering in the 2DEG by assuming
scattering after time of flights which obey a Poisson distribution
with a time constant of $\tau_q$. The electrons in a 2DEG in a
modulation-doped Ga[Al]As heterostructure are predominantly
scattered at the ionized donors in the doping layer, and the
corresponding \emph{screened Coulomb scattering} is known to form an
approximately Gaussian distribution of scattering
angles.\cite{Harrang1985} We therefore assume a Gaussian
distribution of scattering angles $\theta$ with a standard deviation
of $0.1\pi$, centered at $\theta=0$ and limited to $|\theta|
\leq\pi$. Within our model, the times of flight between two
subsequent scattering events form a Poisson distribution with an
expectation value of $\tau $. At a scattering event, the angles
between the initial and the final electron velocity vector direction
are changed according to the distribution function described above.
These two distributions reproduce the experimentally
determined values for $\tau_q$ and $\tau_D$.\\
We remark that our simulation results are rather insensitive to the
chosen distribution function of scattering angles. We have also used
a rectangular distribution, namely a constant probability for
scattering angles $\theta\leq 0.06\pi$ and a probability of zero for
larger angles. Even though we find slightly higher values for the
resistance (about 2.5 \%) at small fields, , the same values are
found in the closed regime.

For each magnetic field, 40000 electrons of Fermi energy $E_F$ are
injected. The transmission is determined by

\begin{equation}
T(E_F,B)=\frac{1}{2W}\int\limits_{-W/2}^{W/2}dy\int\limits_
{-\frac{\pi}{2}}^{\frac{\pi}{2}}cos(\alpha)t(E_F,\alpha, y,B)
d\alpha \label{eq3}
\end{equation}

Here, $\alpha$ is the angle between the x-direction and the
direction in which the electron is injected and $t(E_F,\alpha,y,B)$
is either $1$ or $0$ depending on whether the electron with the
corresponding initial conditions is transmitted or not. We note that
the carrier density under the Dy film may differ from that
underneath the Cr/Au gate, as a consequence of different Schottky
barriers. Hall measurements at voltage probes 1 and 5, as well as at
3 and 7, respectively, indicate that the electron density under the
Dy is roughly 5\% larger as compared with the density measured
outside, but this value is ambiguous since the z-component of the Dy
fringe field superimposes on the homogeneous magnetic field in the
z-direction and we were unable to separate these two contributions
to the modified Hall resistance.\cite{Vancura2000} We have neglected
this density step in our simulations.

The conductance is given by

\begin{equation}
G(E_F,B)=N(E_F)\frac{2e^2}{h}\frac{T(E_F,B)}{1-T(E_F,B)} \label{eq4}
\end{equation}

where $N=2W/\lambda_F=918$ is the number of modes in our Hall bar
and $\lambda_F$ is the Fermi wavelength. We note that according to
eq. \ref{eq4}, the contact resistance ($R_c=h/2e^2N=14.1 \Omega$)
between an infinitely extended 2DEG and the Hall bar does not
contribute to $R_{xx}$.\cite{Beenakker1989,Vancura2000} From eq.
\eqref{eq4}, the longitudinal resistance $R_{xx}$ across the barrier
for a given carrier density is readily obtained from
$R_{xx}(B)\equiv 1/G(E_F,B)$. As a test simulation, we have turned
off all scattering and set the electric field at the edges to zero,
thereby simulating a ballistic magnetic barrier which extends to
infinity in the y-direction. In this case, the numerical results
closely match the corresponding analytical expression
\cite{Kubrak2001} and reproduce the critical angle of
incidence for which the magnetic barrier closes to an accuracy of 1 degree.\\

\begin{figure}
\includegraphics[scale=0.4]{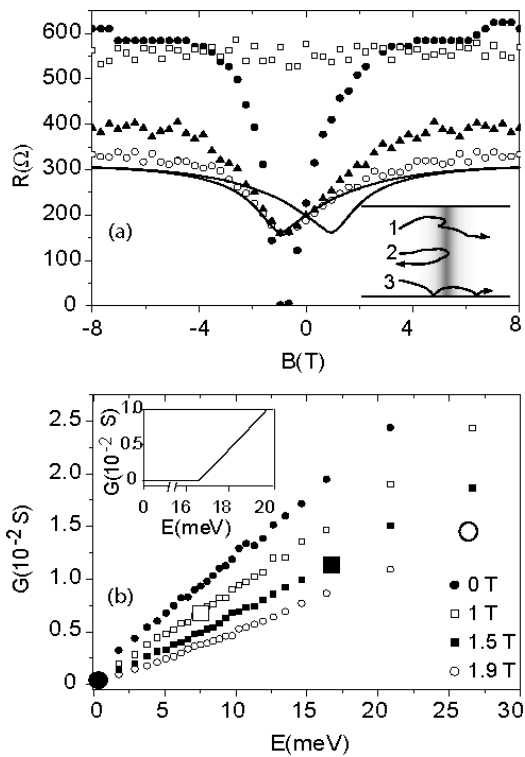}
\caption{(a) Comparison of  model calculations for $R_{xx}(B)$ with
the experimental trace of Fig. 2 (full line), corrected for the
magnetoresistivity of the 2DEG which is set to its constant value at
$B=0\,\mathrm{T}$. Full circles: $R_{xx}$ for a barrier with edges
(edge electric field $|\vec{E}|=10^6\,\mathrm{V/m}$) and no
scattering; triangles: $R_{xx}$ for the barrier with no edges but
scattering according to the experimentally determined scattering
times; open circles: barrier resistance with both scattering and
edge electric field; open squares: $R_{xx}$ for the structure with
edges and a small quantum scattering time of
$\tau_q=0.2\,\mathrm{ps}$. Inset: typical calculated trajectories in
the closed regime: scattering inside the magnetic barrier (trace 1)
as well as $E\times B$ - drifts by edge electric fields (trace 3)
are responsible for a finite resistance. (b) Conductance of the
magnetic barrier as a function of electron energy for several values
of $\mu_0 M$ including edges and scattering. Large symbols in the
figure correspond to the critical energy for which an infinitely
extended barrier in a ballistic system will close (the corresponding
energy dependent conductance for $\mu_0M =1.5\,\mathrm{T}$ is shown
in the inset). } \label{mb4}
\end{figure}

For a comparison of the simulations with the experiments, the
magnetization trace $M(B)$ shown in the inset of Fig. \ref{mb2} is
used to map the height of the magnetic barrier onto the experimental
value of $B$.  Results of the simulations are represented in Fig.
\ref{mb4}(a). Most significantly, the addition of the edge electric
fields to the ballistic system limits the resistance in the closed
regime to finite values. Some electrons that would be reflected at
the barrier away from the edges are pulled through the barrier at
the edges due to the $E \times B$ drift, see the inset in Fig.
\ref{mb4}(a). This effect is the sole reason for a finite resistance
in the closed regime, as long as scattering is disregarded. The
barrier resistance decreases as the edge electric field is
increased. In our simulations, we have assumed an electric field of
$|\vec{E}|=10^6\,\mathrm{V/m}$. This value can be considered an
upper limit, based upon measurements of the steepness of the
confining walls \cite{Fuhrer2001} and in agreement with the
consideration that a potential change of the order of $E_F/e$ cannot
occur over a length smaller than the screening length.
\cite{Larkin1995} We note that the simulated barrier resistance is
only weakly dependent
on the strength of the edge electric field.\\

For an infinitely extended closed magnetic barrier in a disordered
system, the resistance is also kept finite by scattering events in
the barrier region, see Fig. \ref{mb4}(a). This is again illustrated
by the characteristic trajectories, see the inset in Fig.
\ref{mb4}(a): a scattering event may redirect an electron which in
the absence of scattering would be rejected by the barrier. For
elastic mean free paths comparable to or below the spatial extension
of the barrier, the barrier becomes unimportant (open squares in
Fig. \ref{mb4}(a), where the simulated elastic mean free
path was $440\,\mathrm{nm}$).\\

In comparison with the experimental data in Fig. \ref{mb2} (a), we
observe that the numerical trace for $R_{xx}(B)$ that takes only the
scattering into account but disregards edge effects disagrees
significantly with the experiment. Here, the measured background
magnetoresistance (Fig. \ref{mb2}) has been replaced by its value at
$B=0\,\mathrm{T}$. The shape of the measurement trace is reproduced,
but its absolute value differs by up to 30\%. We point out that in
this simulation, the only adjustable parameter is the distribution
of scattering angles under the constraints set by the measured
values for $\tau_D$ and $\tau_q$, which is determined by the details
of the disorder potential landscape.

From this separate discussion of the two mechanisms, it emerges that
a combination of both edge transmission and scattering-induced
transmission determines $R_{xx}(B)$. In fact, inclusion of both
elastic scattering in accordance with the measured scattering times,
as well as an electrostatic edge field of $|\vec{E}|=
10^6\,\mathrm{V/m}$, gives a very good reproduction of the measured
trace, see Fig. \ref{mb4} (a). We refrain from fitting the
experimental data since further uncertainties may have an influence
on this level of accuracy. First, there is a slight asymmetry of the
measured traces, which we attribute to asymmetries in the voltage
probe geometry.\cite{Ibrahim1997} Second, the shape of the magnetic
barrier may deviate from eq. \eqref{eq1}, and it may be
inhomogeneous in the y-direction. Furthermore, the electron density
below the Dy deviates from that below the Cr. This effect could in
principle be avoided by preparing a thin, homogeneous metal
electrode between the semiconductor surface and the ferromagnet.\\

In order to investigate the influence of thermal smearing, we have
calculated the energy dependent conductance $G(E,B)$ by varying the
energy in eq. \eqref{eq4}, from which the conductance at non-zero
temperatures is obtained via

\begin{equation}
G(B,T)=\int\limits_{0}^{\infty}G(E,B)(-\frac{\partial f}{\partial
E})dE \label{eq5}
\end{equation}

where $f$ denotes the Fermi-Dirac distribution function. Fig.
\ref{mb4} (b) shows that the simulated $G(E,B)$ is a nearly linear
function of the electron energy. According to eq. \eqref{eq5},
$G(B,T)$ becomes independent of temperature for $G(E,B)\propto E$. A
similar relation is also approximately found within an analytic
treatment of the infinitely extended, open magnetic barrier in a
ballistic 2DEG \cite{Kubrak2001}, see the inset in Fig. \ref{mb4}
(b). Hence, the inclusion of both edge effects and scattering does
not change this insensitivity of the magnetic barrier resistance to
thermal smearing. We conclude, therefore, that the changes of
$R_{xx}(B)$ with temperature are, besides the temperature dependence
of $\mu_0M(B)$, mainly due to the temperature dependence of the
scattering times. In our experiments, we find that both scattering
times are constant up to $6\,\mathrm{K}$, and in addition, we do not
see significant changes in $R_{xx}(B)$, see Fig. \ref{mb3}, while
for larger temperatures, the observed Shubnikov-de Haas oscillations
no longer allow a meaningful determination of $\tau_q$. Hence, a
reasonable approximation of the measurements at higher temperatures
requires more detailed information regarding the scattering times
and the distribution of scattering angles than available from our
experiments.

We have furthermore studied numerically the effect of Zeeman
splitting on $R_{xx}(B)$ (using an effective g-factor of -0.44) due
to which the two spin directions acquire different Fermi energies
and therefore different partial conductances, resulting in a spin
polarization of the current. Our simulations suggest that the
influence of the spin splitting on $R_{xx}(B)$ is marginal. Also, it
is found numerically that the spin polarization of the current
increases with increasing barrier height in the closed regime, but
is below $10^{-3}$ for all magnetic fields. However, this value is
about a factor of 5 larger than the simulated values for magnetic
barriers without edges, and we conclude that edge transmission tends
to increase the spin polarization. Even accounting for this
increase, the effect for a 2DEG in Ga[Al]As remains very small.
Finally, we have also incorporated magnetic mass effects induced by
the strong parallel magnetic field \cite{Bhattacharya1982} and find
that they are negligible in our parameter range.

\section{\label{sec:5}SUMMARY AND CONCLUSION}
We have studied the resistance of magnetic barriers defined in
Ga[Al]As heterostructures in the quasi-ballistic regime as a
function of in-plane magnetic fields. We have also described the
system numerically within a semiclassical model, and we find that
the finite resistance observed in the closed regime originates from
both elastic scattering in the barrier region and from transmission
via $E\times B$ drifts at the edges of the Hall bar. By using the
scattering times as extracted from the experiment, a very good
agreement between measurement and simulation is obtained, especially
given the uncertainties involved regarding the exact shape and
homogeneity of the magnetic barrier. Furthermore, the barrier
magnetoresistance is insensitive to thermal smearing, spin
polarization and magnetic mass effects. The results also show how
the resistance change induced by the magnetic barrier can be
increased, which may be of importance if one wishes to observe
quantum effects such as resonant tunneling \cite{Matulis1994} or
spin polarization. First of all, both larger mobilities and Hall
bars of reduced width will reduce the scattering in the barrier
region and thereby increase the barrier resistance. Also, defining
soft edges reduces in principle the transmission via $E\times B$
drift effects; our simulations however suggest that very soft edges
with edge fields in the range of $100\,\mathrm{V/m}$ are required to
obtain a noticeable effect. Finally, the deposition of clean
ferromagnetic films under ultra high vacuum conditions should
enhance the saturation magnetization almost up to a factor of two in
our samples.

Our model can easily be extended to describe more complicated
magnetic barrier structures, for example those suggested recently
for use as tunable spin
filters.\cite{Guo2000,Xu2001,Guo2002,Jiang2002,
Zhai2005,Zhai2006}\\

The authors acknowledge the careful reading of the manuscript by
Matthew Jenkins, stimulating discussions with Hengyi Xu as well as
financial support by the \emph{Heinrich-Heine-Universit\"at
D\"usseldorf} .


\begin{thebibliography}{41}
\expandafter\ifx\csname
natexlab\endcsname\relax\def\natexlab#1{#1}\fi
\expandafter\ifx\csname bibnamefont\endcsname\relax
  \def\bibnamefont#1{#1}\fi
\expandafter\ifx\csname bibfnamefont\endcsname\relax
  \def\bibfnamefont#1{#1}\fi
\expandafter\ifx\csname citenamefont\endcsname\relax
  \def\citenamefont#1{#1}\fi
\expandafter\ifx\csname url\endcsname\relax
  \def\url#1{\texttt{#1}}\fi
\expandafter\ifx\csname urlprefix\endcsname\relax\def\urlprefix{URL
}\fi \providecommand{\bibinfo}[2]{#2}
\providecommand{\eprint}[2][]{\url{#2}}

\bibitem[{\citenamefont{Ye et~al.}(1995)\citenamefont{Ye, Weiss, Gerhardts,
  Seeger, von Klitzing, Eberl, and Nickel}}]{Ye1995}
\bibinfo{author}{\bibfnamefont{P.~D.} \bibnamefont{Ye}},
  \bibinfo{author}{\bibfnamefont{D.}~\bibnamefont{Weiss}},
  \bibinfo{author}{\bibfnamefont{R.~R.} \bibnamefont{Gerhardts}},
  \bibinfo{author}{\bibfnamefont{M.}~\bibnamefont{Seeger}},
  \bibinfo{author}{\bibfnamefont{K.}~\bibnamefont{von Klitzing}},
  \bibinfo{author}{\bibfnamefont{K.}~\bibnamefont{Eberl}}, \bibnamefont{and}
  \bibinfo{author}{\bibfnamefont{H.}~\bibnamefont{Nickel}},
  \bibinfo{journal}{Phys. Rev. Lett.} \textbf{\bibinfo{volume}{74}},
  \bibinfo{pages}{3013} (\bibinfo{year}{1995}).

\bibitem[{\citenamefont{Nogaret et~al.}(2003)\citenamefont{Nogaret, Lawton,
  Maude, Portal, and Henini}}]{Nogaret2003}
\bibinfo{author}{\bibfnamefont{A.}~\bibnamefont{Nogaret}},
  \bibinfo{author}{\bibfnamefont{D.~N.} \bibnamefont{Lawton}},
  \bibinfo{author}{\bibfnamefont{D.~K.} \bibnamefont{Maude}},
  \bibinfo{author}{\bibfnamefont{J.~C.} \bibnamefont{Portal}},
  \bibnamefont{and} \bibinfo{author}{\bibfnamefont{M.}~\bibnamefont{Henini}},
  \bibinfo{journal}{Phys. Rev. B} \textbf{\bibinfo{volume}{67}},
  \bibinfo{pages}{165317} (\bibinfo{year}{2003}).

\bibitem[{\citenamefont{Peeters and Li}(1998)}]{Peeters1998}
\bibinfo{author}{\bibfnamefont{F.~M.} \bibnamefont{Peeters}} \bibnamefont{and}
  \bibinfo{author}{\bibfnamefont{X.~Q.} \bibnamefont{Li}},
  \bibinfo{journal}{Appl. Phys. Lett.} \textbf{\bibinfo{volume}{72}},
  \bibinfo{pages}{572} (\bibinfo{year}{1998}).

\bibitem[{\citenamefont{Novoselov et~al.}(2002)\citenamefont{Novoselov, Geim,
  Dubonos, Cornelissens, Peeters, and Maan}}]{Novoselov2002}
\bibinfo{author}{\bibfnamefont{K.~S.} \bibnamefont{Novoselov}},
  \bibinfo{author}{\bibfnamefont{A.~K.} \bibnamefont{Geim}},
  \bibinfo{author}{\bibfnamefont{S.~V.} \bibnamefont{Dubonos}},
  \bibinfo{author}{\bibfnamefont{Y.~G.} \bibnamefont{Cornelissens}},
  \bibinfo{author}{\bibfnamefont{F.~M.} \bibnamefont{Peeters}},
  \bibnamefont{and} \bibinfo{author}{\bibfnamefont{J.~C.} \bibnamefont{Maan}},
  \bibinfo{journal}{Phys. Rev. B} \textbf{\bibinfo{volume}{65}},
  \bibinfo{pages}{233312} (\bibinfo{year}{2002}).

\bibitem[{\citenamefont{Peeters and Matulis}(1993)}]{Peeters1993}
\bibinfo{author}{\bibfnamefont{F.~M.} \bibnamefont{Peeters}} \bibnamefont{and}
  \bibinfo{author}{\bibfnamefont{A.}~\bibnamefont{Matulis}},
  \bibinfo{journal}{Phys. Rev. B} \textbf{\bibinfo{volume}{48}},
  \bibinfo{pages}{15166} (\bibinfo{year}{1993}).

\bibitem[{\citenamefont{Matulis et~al.}(1994)\citenamefont{Matulis, Peeters,
  and Vasilopoulos}}]{Matulis1994}
\bibinfo{author}{\bibfnamefont{A.}~\bibnamefont{Matulis}},
  \bibinfo{author}{\bibfnamefont{F.~M.} \bibnamefont{Peeters}},
  \bibnamefont{and}
  \bibinfo{author}{\bibfnamefont{P.}~\bibnamefont{Vasilopoulos}},
  \bibinfo{journal}{Phys. Rev. Lett.} \textbf{\bibinfo{volume}{72}},
  \bibinfo{pages}{1518} (\bibinfo{year}{1994}).

\bibitem[{\citenamefont{Majumdar}(1996)}]{Majumdar1996}
\bibinfo{author}{\bibfnamefont{A.}~\bibnamefont{Majumdar}},
  \bibinfo{journal}{Phys. Rev. B} \textbf{\bibinfo{volume}{54}},
  \bibinfo{pages}{11911} (\bibinfo{year}{1996}).

\bibitem[{\citenamefont{Guo et~al.}(2000)\citenamefont{Guo, Gu, Zheng, Yu, and
  Kawazoe}}]{Guo2000}
\bibinfo{author}{\bibfnamefont{Y.}~\bibnamefont{Guo}},
  \bibinfo{author}{\bibfnamefont{B.~L.} \bibnamefont{Gu}},
  \bibinfo{author}{\bibfnamefont{Z.}~\bibnamefont{Zheng}},
  \bibinfo{author}{\bibfnamefont{J.~Z.} \bibnamefont{Yu}}, \bibnamefont{and}
  \bibinfo{author}{\bibfnamefont{Y.}~\bibnamefont{Kawazoe}},
  \bibinfo{journal}{Phys. Rev. B} \textbf{\bibinfo{volume}{62}},
  \bibinfo{pages}{2635} (\bibinfo{year}{2000}).

\bibitem[{\citenamefont{Papp and Peeters}(2001{\natexlab{a}})}]{Papp2001a}
\bibinfo{author}{\bibfnamefont{G.}~\bibnamefont{Papp}} \bibnamefont{and}
  \bibinfo{author}{\bibfnamefont{F.~M.} \bibnamefont{Peeters}},
  \bibinfo{journal}{Appl. Phys. Lett.} \textbf{\bibinfo{volume}{78}},
  \bibinfo{pages}{2184} (\bibinfo{year}{2001}{\natexlab{a}}).

\bibitem[{\citenamefont{Papp and Peeters}(2001{\natexlab{b}})}]{Papp2001b}
\bibinfo{author}{\bibfnamefont{G.}~\bibnamefont{Papp}} \bibnamefont{and}
  \bibinfo{author}{\bibfnamefont{F.~M.} \bibnamefont{Peeters}},
  \bibinfo{journal}{Appl. Phys. Lett.} \textbf{\bibinfo{volume}{79}},
  \bibinfo{pages}{3198} (\bibinfo{year}{2001}{\natexlab{b}}).

\bibitem[{\citenamefont{Xu and Okada}(2001)}]{Xu2001}
\bibinfo{author}{\bibfnamefont{H.~Z.} \bibnamefont{Xu}} \bibnamefont{and}
  \bibinfo{author}{\bibfnamefont{Y.}~\bibnamefont{Okada}},
  \bibinfo{journal}{Appl. Phys. Lett.} \textbf{\bibinfo{volume}{79}},
  \bibinfo{pages}{3119} (\bibinfo{year}{2001}).

\bibitem[{\citenamefont{Guo et~al.}(2002)\citenamefont{Guo, Zhai, Gu, and
  Kawazoe}}]{Guo2002}
\bibinfo{author}{\bibfnamefont{Y.}~\bibnamefont{Guo}},
  \bibinfo{author}{\bibfnamefont{F.}~\bibnamefont{Zhai}},
  \bibinfo{author}{\bibfnamefont{B.~L.} \bibnamefont{Gu}}, \bibnamefont{and}
  \bibinfo{author}{\bibfnamefont{Y.}~\bibnamefont{Kawazoe}},
  \bibinfo{journal}{Phys. Rev. B} \textbf{\bibinfo{volume}{66}},
  \bibinfo{pages}{045312} (\bibinfo{year}{2002}).

\bibitem[{\citenamefont{Jiang et~al.}(2002)\citenamefont{Jiang, Jalil, and
  Low}}]{Jiang2002}
\bibinfo{author}{\bibfnamefont{Y.}~\bibnamefont{Jiang}},
  \bibinfo{author}{\bibfnamefont{M.~B.~A.} \bibnamefont{Jalil}},
  \bibnamefont{and} \bibinfo{author}{\bibfnamefont{T.}~\bibnamefont{Low}},
  \bibinfo{journal}{Appl. Phys. Lett.} \textbf{\bibinfo{volume}{80}},
  \bibinfo{pages}{1673} (\bibinfo{year}{2002}).

\bibitem[{\citenamefont{Zhai and Xu}(2005)}]{Zhai2005}
\bibinfo{author}{\bibfnamefont{F.}~\bibnamefont{Zhai}} \bibnamefont{and}
  \bibinfo{author}{\bibfnamefont{H.~Q.} \bibnamefont{Xu}},
  \bibinfo{journal}{Phys. Rev. B} \textbf{\bibinfo{volume}{72}},
  \bibinfo{pages}{085314} (\bibinfo{year}{2005}).

\bibitem[{\citenamefont{Zhai and Xu}(2006)}]{Zhai2006}
\bibinfo{author}{\bibfnamefont{F.}~\bibnamefont{Zhai}} \bibnamefont{and}
  \bibinfo{author}{\bibfnamefont{H.~Q.} \bibnamefont{Xu}},
  \bibinfo{journal}{Appl. Phys. Lett.} \textbf{\bibinfo{volume}{88}},
  \bibinfo{pages}{032502} (\bibinfo{year}{2006}).

\bibitem[{\citenamefont{Leadbeater et~al.}(1995)\citenamefont{Leadbeater,
  Foden, Burroughes, Pepper, Burke, Wang, Grimshaw, and
  Ritchie}}]{Leadbeater1995}
\bibinfo{author}{\bibfnamefont{M.~L.} \bibnamefont{Leadbeater}},
  \bibinfo{author}{\bibfnamefont{C.~L.} \bibnamefont{Foden}},
  \bibinfo{author}{\bibfnamefont{J.~H.} \bibnamefont{Burroughes}},
  \bibinfo{author}{\bibfnamefont{M.}~\bibnamefont{Pepper}},
  \bibinfo{author}{\bibfnamefont{T.~M.} \bibnamefont{Burke}},
  \bibinfo{author}{\bibfnamefont{L.~L.} \bibnamefont{Wang}},
  \bibinfo{author}{\bibfnamefont{M.~P.} \bibnamefont{Grimshaw}},
  \bibnamefont{and} \bibinfo{author}{\bibfnamefont{D.~A.}
  \bibnamefont{Ritchie}}, \bibinfo{journal}{Phys. Rev. B}
  \textbf{\bibinfo{volume}{52}}, \bibinfo{pages}{R8629} (\bibinfo{year}{1995}).

\bibitem[{\citenamefont{Ibrahim et~al.}(1997)\citenamefont{Ibrahim, Schweikert,
  and Peeters}}]{Ibrahim1997}
\bibinfo{author}{\bibfnamefont{I.~S.} \bibnamefont{Ibrahim}},
  \bibinfo{author}{\bibfnamefont{V.~A.} \bibnamefont{Schweikert}},
  \bibnamefont{and} \bibinfo{author}{\bibfnamefont{F.~M.}
  \bibnamefont{Peeters}}, \bibinfo{journal}{Phys. Rev. B}
  \textbf{\bibinfo{volume}{56}}, \bibinfo{pages}{7508} (\bibinfo{year}{1997}).

\bibitem[{\citenamefont{Monzon et~al.}(1997)\citenamefont{Monzon, Johnson, and
  Roukes}}]{Monzon1997}
\bibinfo{author}{\bibfnamefont{F.~G.} \bibnamefont{Monzon}},
  \bibinfo{author}{\bibfnamefont{M.}~\bibnamefont{Johnson}}, \bibnamefont{and}
  \bibinfo{author}{\bibfnamefont{M.~L.} \bibnamefont{Roukes}},
  \bibinfo{journal}{Appl. Phys. Lett.} \textbf{\bibinfo{volume}{71}},
  \bibinfo{pages}{3087} (\bibinfo{year}{1997}).

\bibitem[{\citenamefont{Kubrak et~al.}(2000)\citenamefont{Kubrak, Neumann,
  Gallagher, Main, and Henini}}]{Kubrak2000}
\bibinfo{author}{\bibfnamefont{V.}~\bibnamefont{Kubrak}},
  \bibinfo{author}{\bibfnamefont{A.~C.} \bibnamefont{Neumann}},
  \bibinfo{author}{\bibfnamefont{B.~L.} \bibnamefont{Gallagher}},
  \bibinfo{author}{\bibfnamefont{P.~C.} \bibnamefont{Main}}, \bibnamefont{and}
  \bibinfo{author}{\bibfnamefont{M.}~\bibnamefont{Henini}},
  \bibinfo{journal}{J. Appl. Phys.} \textbf{\bibinfo{volume}{87}},
  \bibinfo{pages}{5986} (\bibinfo{year}{2000}).

\bibitem[{\citenamefont{Van$\mathrm{\check{c}}$ura
  et~al.}(2000)\citenamefont{Van$\mathrm{\check{c}}$ura, Ihn, Broderick,
  Ensslin, Wegscheider, and Bichler}}]{Vancura2000}
\bibinfo{author}{\bibfnamefont{T.}~\bibnamefont{Van$\mathrm{\check{c}}$ura}},
  \bibinfo{author}{\bibfnamefont{T.}~\bibnamefont{Ihn}},
  \bibinfo{author}{\bibfnamefont{S.}~\bibnamefont{Broderick}},
  \bibinfo{author}{\bibfnamefont{K.}~\bibnamefont{Ensslin}},
  \bibinfo{author}{\bibfnamefont{W.}~\bibnamefont{Wegscheider}},
  \bibnamefont{and} \bibinfo{author}{\bibfnamefont{M.}~\bibnamefont{Bichler}},
  \bibinfo{journal}{Phys. Rev. B} \textbf{\bibinfo{volume}{62}},
  \bibinfo{pages}{5074} (\bibinfo{year}{2000}).

\bibitem[{\citenamefont{Kubrak et~al.}(2001)\citenamefont{Kubrak, Edmonds,
  Neumann, Gallagher, Main, Henini, Marrows, Hickey, and Thoms}}]{Kubrak2001}
\bibinfo{author}{\bibfnamefont{V.}~\bibnamefont{Kubrak}},
  \bibinfo{author}{\bibfnamefont{K.~W.} \bibnamefont{Edmonds}},
  \bibinfo{author}{\bibfnamefont{A.~C.} \bibnamefont{Neumann}},
  \bibinfo{author}{\bibfnamefont{B.~L.} \bibnamefont{Gallagher}},
  \bibinfo{author}{\bibfnamefont{P.~C.} \bibnamefont{Main}},
  \bibinfo{author}{\bibfnamefont{M.}~\bibnamefont{Henini}},
  \bibinfo{author}{\bibfnamefont{C.~H.} \bibnamefont{Marrows}},
  \bibinfo{author}{\bibfnamefont{B.~J.} \bibnamefont{Hickey}},
  \bibnamefont{and} \bibinfo{author}{\bibfnamefont{S.}~\bibnamefont{Thoms}},
  \bibinfo{journal}{IEEE Trans. Magn.} \textbf{\bibinfo{volume}{37}},
  \bibinfo{pages}{1992} (\bibinfo{year}{2001}).

\bibitem[{\citenamefont{Gallagher et~al.}(2001)\citenamefont{Gallagher, Kubrak,
  Rushforth, Neumann, Edmonds, Main, Henini, Marrows, Hickey, and
  Thoms}}]{Gallagher2001}
\bibinfo{author}{\bibfnamefont{B.~L.} \bibnamefont{Gallagher}},
  \bibinfo{author}{\bibfnamefont{V.}~\bibnamefont{Kubrak}},
  \bibinfo{author}{\bibfnamefont{A.~W.} \bibnamefont{Rushforth}},
  \bibinfo{author}{\bibfnamefont{A.~C.} \bibnamefont{Neumann}},
  \bibinfo{author}{\bibfnamefont{K.~W.} \bibnamefont{Edmonds}},
  \bibinfo{author}{\bibfnamefont{P.~C.} \bibnamefont{Main}},
  \bibinfo{author}{\bibfnamefont{M.}~\bibnamefont{Henini}},
  \bibinfo{author}{\bibfnamefont{C.~H.} \bibnamefont{Marrows}},
  \bibinfo{author}{\bibfnamefont{B.~J.} \bibnamefont{Hickey}},
  \bibnamefont{and} \bibinfo{author}{\bibfnamefont{S.}~\bibnamefont{Thoms}},
  \bibinfo{journal}{Physica E} \textbf{\bibinfo{volume}{11}},
  \bibinfo{pages}{171} (\bibinfo{year}{2001}).

\bibitem[{\citenamefont{Hong et~al.}(2002)\citenamefont{Hong, Kubrak, Edmonds,
  Neumann, Gallagher, Main, Henini, Marrows, Hickey, and Thoms}}]{Hong2002}
\bibinfo{author}{\bibfnamefont{J.}~\bibnamefont{Hong}},
  \bibinfo{author}{\bibfnamefont{V.}~\bibnamefont{Kubrak}},
  \bibinfo{author}{\bibfnamefont{K.~W.} \bibnamefont{Edmonds}},
  \bibinfo{author}{\bibfnamefont{A.~C.} \bibnamefont{Neumann}},
  \bibinfo{author}{\bibfnamefont{B.~L.} \bibnamefont{Gallagher}},
  \bibinfo{author}{\bibfnamefont{P.~C.} \bibnamefont{Main}},
  \bibinfo{author}{\bibfnamefont{M.}~\bibnamefont{Henini}},
  \bibinfo{author}{\bibfnamefont{C.~H.} \bibnamefont{Marrows}},
  \bibinfo{author}{\bibfnamefont{B.~J.} \bibnamefont{Hickey}},
  \bibnamefont{and} \bibinfo{author}{\bibfnamefont{S.}~\bibnamefont{Thoms}},
  \bibinfo{journal}{Physica E} \textbf{\bibinfo{volume}{12}},
  \bibinfo{pages}{229} (\bibinfo{year}{2002}).

\bibitem[{\citenamefont{Lu et~al.}(2002)\citenamefont{Lu, Zhang, and
  Yan}}]{Lu2002}
\bibinfo{author}{\bibfnamefont{M.-W.} \bibnamefont{Lu}},
  \bibinfo{author}{\bibfnamefont{L.-D.} \bibnamefont{Zhang}}, \bibnamefont{and}
  \bibinfo{author}{\bibfnamefont{X.-H.} \bibnamefont{Yan}},
  \bibinfo{journal}{Phys. Rev. B} \textbf{\bibinfo{volume}{66}},
  \bibinfo{pages}{224412} (\bibinfo{year}{2002}).

\bibitem[{\citenamefont{Xu and Guo}(2005)}]{Xu2005}
\bibinfo{author}{\bibfnamefont{W.}~\bibnamefont{Xu}} \bibnamefont{and}
  \bibinfo{author}{\bibfnamefont{Y.}~\bibnamefont{Guo}},
  \bibinfo{journal}{Phys. Lett. A} \textbf{\bibinfo{volume}{340}},
  \bibinfo{pages}{281} (\bibinfo{year}{2005}).

\bibitem[{\citenamefont{Behrendt et~al.}(1958)\citenamefont{Behrendt, Legvold,
  and Spedding}}]{Behrendt1958}
\bibinfo{author}{\bibfnamefont{D.~R.} \bibnamefont{Behrendt}},
  \bibinfo{author}{\bibfnamefont{S.}~\bibnamefont{Legvold}}, \bibnamefont{and}
  \bibinfo{author}{\bibfnamefont{F.~H.} \bibnamefont{Spedding}},
  \bibinfo{journal}{Phys. Rev.} \textbf{\bibinfo{volume}{109}},
  \bibinfo{pages}{1544} (\bibinfo{year}{1958}).

\bibitem[{\citenamefont{Stepankin}(1995)}]{Stepankin1995}
\bibinfo{author}{\bibfnamefont{V.}~\bibnamefont{Stepankin}},
  \bibinfo{journal}{Physica B} \textbf{\bibinfo{volume}{211}},
  \bibinfo{pages}{345} (\bibinfo{year}{1995}).

\bibitem[{\citenamefont{B$\mathrm{\ddot{u}}$ttiker}(1986)}]{Buttiker1986}
\bibinfo{author}{\bibfnamefont{M.}~\bibnamefont{B$\mathrm{\ddot{u}}$ttiker}},
  \bibinfo{journal}{Phys. Rev. Lett.} \textbf{\bibinfo{volume}{57}},
  \bibinfo{pages}{1761} (\bibinfo{year}{1986}).

\bibitem[{\citenamefont{Beenakker and van Houten}(1989)}]{Beenakker1989}
\bibinfo{author}{\bibfnamefont{C.~W.~J.} \bibnamefont{Beenakker}}
  \bibnamefont{and} \bibinfo{author}{\bibfnamefont{H.}~\bibnamefont{van
  Houten}}, \bibinfo{journal}{Phys. Rev. Lett.} \textbf{\bibinfo{volume}{63}},
  \bibinfo{pages}{1857} (\bibinfo{year}{1989}).

\bibitem[{Int()}]{IntelliEpi}
\bibinfo{note}{The heterostructures have been purchased from Intelligent
  Epitaxy Tech., Richardson, TX (USA).}

\bibitem[{\citenamefont{Ando et~al.}(1982)\citenamefont{Ando, Fowler, and
  Stern}}]{Ando1982}
\bibinfo{author}{\bibfnamefont{T.}~\bibnamefont{Ando}},
  \bibinfo{author}{\bibfnamefont{A.~B.} \bibnamefont{Fowler}},
  \bibnamefont{and} \bibinfo{author}{\bibfnamefont{F.}~\bibnamefont{Stern}},
  \bibinfo{journal}{Rev. Mod. Phys.} \textbf{\bibinfo{volume}{54}},
  \bibinfo{pages}{437} (\bibinfo{year}{1982}).

\bibitem[{\citenamefont{Lee and Ramakrishnan}(1985)}]{Lee1985}
\bibinfo{author}{\bibfnamefont{P.~A.} \bibnamefont{Lee}} \bibnamefont{and}
  \bibinfo{author}{\bibfnamefont{T.~V.} \bibnamefont{Ramakrishnan}},
  \bibinfo{journal}{Rev. Mod. Phys.} \textbf{\bibinfo{volume}{57}},
  \bibinfo{pages}{287} (\bibinfo{year}{1985}).

\bibitem[{\citenamefont{Bhattacharya}(1982)}]{Bhattacharya1982}
\bibinfo{author}{\bibfnamefont{S.~K.} \bibnamefont{Bhattacharya}},
  \bibinfo{journal}{Phys. Rev. B} \textbf{\bibinfo{volume}{25}},
  \bibinfo{pages}{3756} (\bibinfo{year}{1982}).

\bibitem[{\citenamefont{Johnson et~al.}(1997)\citenamefont{Johnson, Bennett,
  Yang, Miller, and Shanabrook}}]{Johnson1997}
\bibinfo{author}{\bibfnamefont{M.}~\bibnamefont{Johnson}},
  \bibinfo{author}{\bibfnamefont{B.~R.} \bibnamefont{Bennett}},
  \bibinfo{author}{\bibfnamefont{M.~J.} \bibnamefont{Yang}},
  \bibinfo{author}{\bibfnamefont{M.~M.} \bibnamefont{Miller}},
  \bibnamefont{and} \bibinfo{author}{\bibfnamefont{B.~V.}
  \bibnamefont{Shanabrook}}, \bibinfo{journal}{Appl. Phys. Lett.}
  \textbf{\bibinfo{volume}{71}}, \bibinfo{pages}{974} (\bibinfo{year}{1997}).

\bibitem[{\citenamefont{Reijniers and Peeters}(1998)}]{Reijniers1998}
\bibinfo{author}{\bibfnamefont{J.}~\bibnamefont{Reijniers}} \bibnamefont{and}
  \bibinfo{author}{\bibfnamefont{F.~M.} \bibnamefont{Peeters}},
  \bibinfo{journal}{Appl. Phys. Lett.} \textbf{\bibinfo{volume}{73}},
  \bibinfo{pages}{357} (\bibinfo{year}{1998}).

\bibitem[{\citenamefont{Reijniers and Peeters}(2000)}]{Reijniers2000}
\bibinfo{author}{\bibfnamefont{J.}~\bibnamefont{Reijniers}} \bibnamefont{and}
  \bibinfo{author}{\bibfnamefont{F.~M.} \bibnamefont{Peeters}},
  \bibinfo{journal}{J. Appl. Phys.} \textbf{\bibinfo{volume}{87}},
  \bibinfo{pages}{8088} (\bibinfo{year}{2000}).

\bibitem[{\citenamefont{Liu et~al.}(1998)\citenamefont{Liu, Guillou, Kent,
  Stupian, and Leung}}]{Liu1998}
\bibinfo{author}{\bibfnamefont{S.}~\bibnamefont{Liu}},
  \bibinfo{author}{\bibfnamefont{H.}~\bibnamefont{Guillou}},
  \bibinfo{author}{\bibfnamefont{A.~D.} \bibnamefont{Kent}},
  \bibinfo{author}{\bibfnamefont{G.~W.} \bibnamefont{Stupian}},
  \bibnamefont{and} \bibinfo{author}{\bibfnamefont{M.~S.} \bibnamefont{Leung}},
  \bibinfo{journal}{J. Appl. Phys.} \textbf{\bibinfo{volume}{83}},
  \bibinfo{pages}{6161} (\bibinfo{year}{1998}).

\bibitem[{\citenamefont{Cornelissens and Peeters}(2002)}]{Cornelissens2002}
\bibinfo{author}{\bibfnamefont{Y.~G.} \bibnamefont{Cornelissens}}
  \bibnamefont{and} \bibinfo{author}{\bibfnamefont{F.~M.}
  \bibnamefont{Peeters}}, \bibinfo{journal}{J. Appl. Phys.}
  \textbf{\bibinfo{volume}{92}}, \bibinfo{pages}{2006} (\bibinfo{year}{2002}).

\bibitem[{\citenamefont{Harrang et~al.}(1985)\citenamefont{Harrang, Higgins,
  Goodall, Jay, Laviron, and Delescluse}}]{Harrang1985}
\bibinfo{author}{\bibfnamefont{J.~P.} \bibnamefont{Harrang}},
  \bibinfo{author}{\bibfnamefont{R.~J.} \bibnamefont{Higgins}},
  \bibinfo{author}{\bibfnamefont{R.~K.} \bibnamefont{Goodall}},
  \bibinfo{author}{\bibfnamefont{P.~R.} \bibnamefont{Jay}},
  \bibinfo{author}{\bibfnamefont{M.}~\bibnamefont{Laviron}}, \bibnamefont{and}
  \bibinfo{author}{\bibfnamefont{P.}~\bibnamefont{Delescluse}},
  \bibinfo{journal}{Phys. Rev. B} \textbf{\bibinfo{volume}{32}},
  \bibinfo{pages}{8126} (\bibinfo{year}{1985}).

\bibitem[{\citenamefont{Fuhrer et~al.}(2001)\citenamefont{Fuhrer,
  L$\mathrm{\ddot{u}}$scher, Heinzel, Ensslin, Wegscheider, and
  Bichler}}]{Fuhrer2001}
\bibinfo{author}{\bibfnamefont{A.}~\bibnamefont{Fuhrer}},
  \bibinfo{author}{\bibfnamefont{S.}~\bibnamefont{L$\mathrm{\ddot{u}}$scher}},
  \bibinfo{author}{\bibfnamefont{T.}~\bibnamefont{Heinzel}},
  \bibinfo{author}{\bibfnamefont{K.}~\bibnamefont{Ensslin}},
  \bibinfo{author}{\bibfnamefont{W.}~\bibnamefont{Wegscheider}},
  \bibnamefont{and} \bibinfo{author}{\bibfnamefont{M.}~\bibnamefont{Bichler}},
  \bibinfo{journal}{Phys. Rev. B} \textbf{\bibinfo{volume}{63}},
  \bibinfo{pages}{125309} (\bibinfo{year}{2001}).

\bibitem[{\citenamefont{Larkin and Davies}(1995)}]{Larkin1995}
\bibinfo{author}{\bibfnamefont{I.~A.} \bibnamefont{Larkin}} \bibnamefont{and}
  \bibinfo{author}{\bibfnamefont{J.~H.} \bibnamefont{Davies}},
  \bibinfo{journal}{Phys. Rev. B} \textbf{\bibinfo{volume}{52}},
  \bibinfo{pages}{5535} (\bibinfo{year}{1995}).

\end{thebibliography}

\end{document}